\documentclass[10pt,twocolumn]{revtex4-1}

\usepackage{amsmath, amsfonts, amssymb}     %
\usepackage{graphicx}                       %
\usepackage{xfrac}                          %
\usepackage{grffile}                        %
\usepackage{color,colortbl}

\usepackage[colorlinks, citecolor=blue, urlcolor=red, linkcolor=blue]{hyperref}

\makeatletter

\def\CT@@do@color{%
  \global\let\CT@do@color\relax
  \@tempdima\wd\z@
  \advance\@tempdima\@tempdimb
  \advance\@tempdima\@tempdimc
  \advance\@tempdimb\tabcolsep
  \advance\@tempdimc\tabcolsep
  \advance\@tempdima2\tabcolsep
  \kern-\@tempdimb
  \leaders\vrule
  \hskip\@tempdima\@plus  1fill
  \kern-\@tempdimc
\hskip-\wd\z@ \@plus -1fill }
\makeatother

\renewcommand{\eqref}{Eq.~\ref}

\newcommand{\gdot}[0] {\dot{\gamma}}
\newcommand{\gdotbar}[0] {\bar{\dot{\gamma}}}
\newcommand{\gdotb}[0] {\bar{\dot{\gamma}}}
\newcommand{\gdotf}[0] {\tilde{\dot{\gamma}}}

\newcommand{\dob}[0] {\Delta_{\hspace{-1pt}\gdot}}
\newcommand{\dobc}[0] {\Delta^{\hspace{-1pt}c}_{\hspace{-1pt}\gdot}}

\newcommand{\xhat}{\hat{\mathbf{x}}}
\newcommand{\yhat}{\hat{\mathbf{y}}}
\newcommand{\zhat}{\hat{\mathbf{z}}}

\newcommand{\vecv}[1]{\mathbf{{#1}}}
\newcommand{\tens}[1]{\mathbf{{#1}}}

\newcommand{\etaa}{\eta_{\rm a}}

\newcommand{\be}{\begin{equation}}
\newcommand{\ee}{\end{equation}}
\newcommand{\bea}{\begin{eqnarray}}
\newcommand{\eea}{\end{eqnarray}}

\begin{document}
\title{Edge-induced shear banding in entangled polymeric fluids}
\author{Ewan J. Hemingway and Suzanne M. Fielding}
\affiliation{Department of Physics, Durham University, Science Laboratories,
  South Road, Durham DH1 3LE, UK}
\date{\today}
\begin{abstract}

Despite decades of research, the question of whether solutions and
melts of highly entangled polymers exhibit shear banding as their
steady state response to a steadily imposed shear flow remains
controversial. From a theoretical viewpoint, an important unanswered
question is whether the underlying constitutive curve of shear stress
$\sigma$ as a function of shear rate $\gdot$ (for states of
homogeneous shear) is monotonic, or has a region of negative slope,
$d\sigma/d\gdot<0$, which would trigger banding. Attempts to settle
the question experimentally via velocimetry of the flow field inside
the fluid are often confounded by an instability of the free surface
where the sample meets the outside air, known as `edge fracture'.
Here we show by numerical simulation that in fact even only very modest edge
disturbances -- which are the precursor of full edge fracture but
might well, in themselves, go unnoticed experimentally -- can cause
strong secondary flows in the form of shear bands that invade deep
into the fluid bulk. Crucially, this is true even when the underlying
constitutive curve is monotonically increasing, precluding true bulk shear
banding in the absence of edge effects.

\end{abstract}
\date{\today}
\maketitle

Polymeric fluids display exotic nonlinear rheological (deformation and
flow) properties, stemming from the complicated underlying dynamics of
their constituent entangled chainlike molecules.  When subject to an
imposed shear, for example, they commonly exhibit a heterogeneous flow
response in which bands of differing shear rate form, with layer
normals in the flow-gradient direction. This phenomenon of `shear
banding' has been widely observed during the transient, time-dependent
process whereby a steady flowing state is established out of an
initial rest state, following the switch on of a
flow~\cite{Ravindranath2008, Boukany2009, Boukany2009a} or
load~\cite{ Hu2010, Boukany2009, Boukany2009a}; and in the perpetually
time-dependent protocol of large amplitude oscillatory
shear~\cite{osc1,osc2}. It has been successfully
captured~\cite{Adams2009,Adams2011,Moorcroft2013,Moorcroft2014,Carter,Ianniruberto2017}
in rheological constitutive models based on molecular
theories~\cite{Likhtman2003,doi1988theory} of polymer dynamics that
posit the dominant mode of stress relaxation to be one of `reptation',
in which any molecule snakes out of an effective tube formed from
entanglements with its neighbours.

Perhaps surprisingly, the more basic question of whether shear bands
form the ultimate {\em steady flowing state} in entangled polymers
remains intensely controversial, despite decades of research. From a
theoretical viewpoint, an important issue concerns whether the
underlying constitutive curve of shear stress $\sigma$ as a function
of shear rate $\gdot$ (for states of stationary homogeneous shear) is
monotonically increasing, or instead has a region of negative slope,
$d\sigma/d\gdot<0$. The latter would necessarily imply homogeneous
shear to be unstable, leading to bulk banding in the steady flowing
state. While the original reptation theory~\cite{doi1988theory}
predicted non-monotonicity, more recent extensions to it incorporating
the additional molecular processes of convective constraint release
and chain stretch
relaxation~\cite{Ianniruberto1996,Ianniruberto2001,Graham2003} can, at
least in principle, restore monotonicity and (in melts) eliminate
steady state banding. (In solutions with a strong enough coupling
between flow and concentration fluctuations, steady state banding has been
predicted to occur even if the constitutive curve is
monotonic~\cite{FO1,FO2,Leal1,Germann}.) Whether they do so in practice,
however, depends on the number of entanglements per molecule and on
the level of convective constraint release, which is a priori unknown.

Just as this debate remains unsettled theoretically, studies aimed at resolving it experimentally have likewise proved controversial. The experimentally measured flow curve $\sigma(\gdotbar)$ is always monotonically increasing, but with a characteristically rather flat region spanning typically 1-4 decades in shear rate~\cite{Sui2007,Hu2007,Hu2010,Li2015,Tapadia2004}, depending on the fluid in question.  Whether the underlying constitutive curve $\sigma(\gdot)$ is itself monotonic is not settled simply by measuring the flow curve, however, because a non-monotonic constitutive curve would lead to shear banding, which restores a monotonic flow curve $\sigma(\gdotbar)$ for the composite banded flow, with $\gdotbar$ the shear rate averaged across the bands. (For homogeneous flow, $\gdotbar=\gdot$ everywhere and the constitutive curve and flow curve coincide.)

The question of whether shear bands are present must therefore instead be
investigated by explicit velocimetric studies of the flow field inside the
sample. Tapadia and Wang ~\cite{Tapadia2006} gave evidence for steady
state shear banding in entangled polymer solutions, suggesting a
non-monotonic underlying constitutive curve.  In contrast, Hu et
al. \cite{Hu2007} observed shear banding only transiently during shear
startup, giving way at longer times to homogeneous shear, suggesting a
monotonic constitutive curve. Later work on more highly entangled
samples did however report long-lived shear bands in some experimental
runs, but not in others~\cite{Hu2010}, even (in some cases) when
repeated for the same imposed loads or flow rates.

While the aim in these velocimetry experiments is to measure the
fluid's true bulk flow behaviour, in practice all rotational shear
rheometers have a free surface where the fluid sample meets the
outside air. Care is obviously then needed to measure the flow field
as far into the sample as possible, at a depth from the surface that
is many multiples of the gap width between the shearing plates. Even
then, however, polymeric fluids are known to be highly susceptible to
`edge fracture'~\cite{Tanner1983,Lee1992,Keentok1999,Hemingway2017},
in which this free surface between the fluid
sample and air destabilises when the fluid is sheared. This can lead
to secondary flows penetrating some depth into the bulk. Indeed, edge
fracture was discussed as a possible source of the variability between
runs mentioned above~\cite{Hu2010}. Significant possible edge fracture
in experiments concerning the presence or absence of shear banding in
entangled polymers has likewise been discussed extensively in
Refs.~\cite{Sui2007,Schweizer2008,Li2013,Wang2014,Li2014a,Li2015,Boukany2015,WangOpenAccess}.

In this Letter, we report for the first time simulations exploring the
complicated dynamical interplay between these surface and bulk
instabilities in sheared polymeric fluids. Our principal contributions
are threefold. First, we show that only modest deformations of the
sample edge -- which are the precursors of true edge fracture but may
(in themselves) go unnoticed -- can indeed lead to secondary flows
that penetrate some distance into the fluid bulk. Second, for a
material with a bulk constitutive curve that is rather flat (but still
monotonically increasing, e.g., comparable to that measured experimentally
in Ref.~\cite{Hu2007}), these secondary flows can be very strong and
can furthermore invade the bulk to up to depths of 10 - 20 gap widths
in from the sample edge, which is the maximum depth typically attained
experimentally due to the finite aspect ratio of any sample. Third,
these secondary flows take the form of shear bands. Importantly, this
is true despite the constitutive curve being monotonic in our
simulations, precluding true bulk shear banding in the absence of any
surface disturbance.
This work therefore shows that only modest precursors of the {\em surface}
transition of edge fracture can precipitate a strong quasi-{\em bulk}
shear banding effect far into the sample.

\begin{figure}[!t]
  \includegraphics[width=\columnwidth]{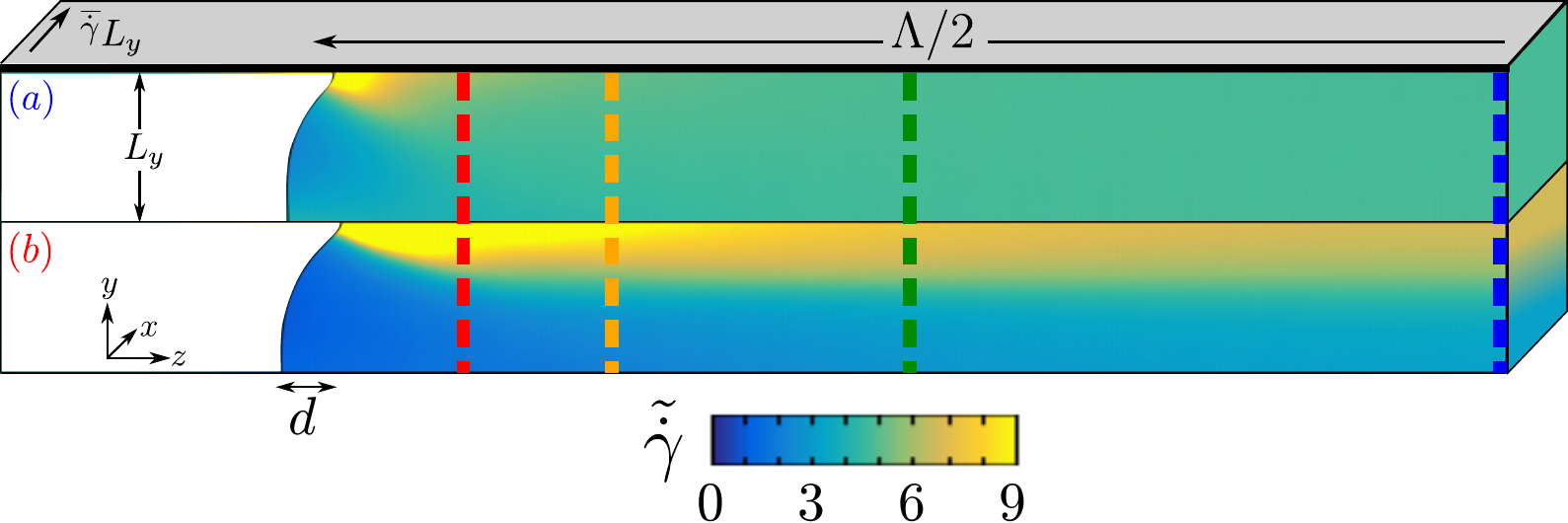}
  \caption{Shear-rate colourmaps in the steady flowing state. {\bf
  Top:} A fluid with the moderately sloping constitutive curve shown
  as (a) in Fig.~\ref{fig:constit} (left) exhibits homogeneous bulk
  flow.  {\bf Bottom:} A fluid with the flatter constitutive curve (b)
  in Fig. ~\ref{fig:constit} (left) shows strong apparent quasi-bulk
  shear banding.  Dashed coloured lines show positions at which the
  velocity profiles of Fig.~\ref{fig:constit} (right) are
  taken. Parameters: $\Lambda = 16.0$, $\gdotb = 4.7$, $\Gamma = 0.16$ with $\eta =
  0.02, 0.006$ giving $n=0.45, 1.06$ (top, bottom).}
  \label{fig:snapshots}
\end{figure}

As shown in Fig.~\ref{fig:snapshots}, we consider a planar slab of
fluid sheared at rate $\gdotb$ between hard walls at $y=0, L_y$.
The flow direction is denoted $\xhat$ and the flow-gradient direction
$\yhat$.  The surfaces of the fluid sample in the vorticity direction
$\zhat$ are in contact with the air. The sample length in that
direction (in the initial unsheared state) is $\Lambda$. Our
simulation box has length $L_z$, with periodic boundary conditions in
$z$. (Only its left half is shown in Fig.~\ref{fig:snapshots}.) At the
plates we impose boundary conditions of no slip and no permeation.
Translational invariance is assumed in $x$.

The total stress $\tens{T}$ in any fluid element is taken to comprise
an isotropic contribution with pressure $p$, a Newtonian contribution
characterised by a viscosity $\eta$, and a slow viscoelastic
contribution $\tens{\Sigma}$ from the polymer chains. The Newtonian
part models contributions from both the background solvent, and also
from fast intrachain polymeric relaxation modes. We assume conditions
of creeping flow, with the force balance condition
$\nabla\cdot\tens{T}=0$. This gives
$\eta\nabla^2\vecv{v}+\nabla.\tens{\Sigma}-\nabla p=0$ inside the
fluid and $\etaa\nabla^2\vecv{v}-\nabla p=0$ in the outside air, with
$\etaa$ the air viscosity. The pressure field $p(\vecv{r},t)$ is
determined by enforcing incompressible flow, such that the velocity
field $\vecv{v}(\vecv{r},t)$ obeys $\nabla\cdot\vecv{v}=0$.  The dynamics
of the viscoelastic stress $\tens{\Sigma}$ is taken to obey the
diffusive Giesekus model~\cite{Giesekus1982}:
\be
D_t\tens{\Sigma} = 2G\tens{D} + \,
\tens{\Sigma}\cdot\nabla\vecv{v}\,+\,\nabla\vecv{v}^{\rm T}\cdot\tens{\Sigma}-\frac{1}{\tau}\left(\tens{\Sigma}+\alpha\tens{\Sigma}^2\right)+D\nabla^2\tens{\Sigma},
\label{eqn:vece}
\ee
in which $D_t\tens{\Sigma}\equiv
\partial_t\tens{\Sigma}+\vecv{v}.\nabla\tens{\Sigma}$ captures Galilean invariance;
$\nabla\vecv{v}_{\alpha\beta}=\partial_\alpha v_\beta$ and
$\tens{D}=\tfrac{1}{2}(\nabla \vecv{v} + \nabla \vecv{v}^T)$.  The
first three terms on the RHS of Eqn.~\ref{eqn:vece} capture the loading of viscoelastic
\begin{figure}
  \includegraphics[width=\columnwidth]{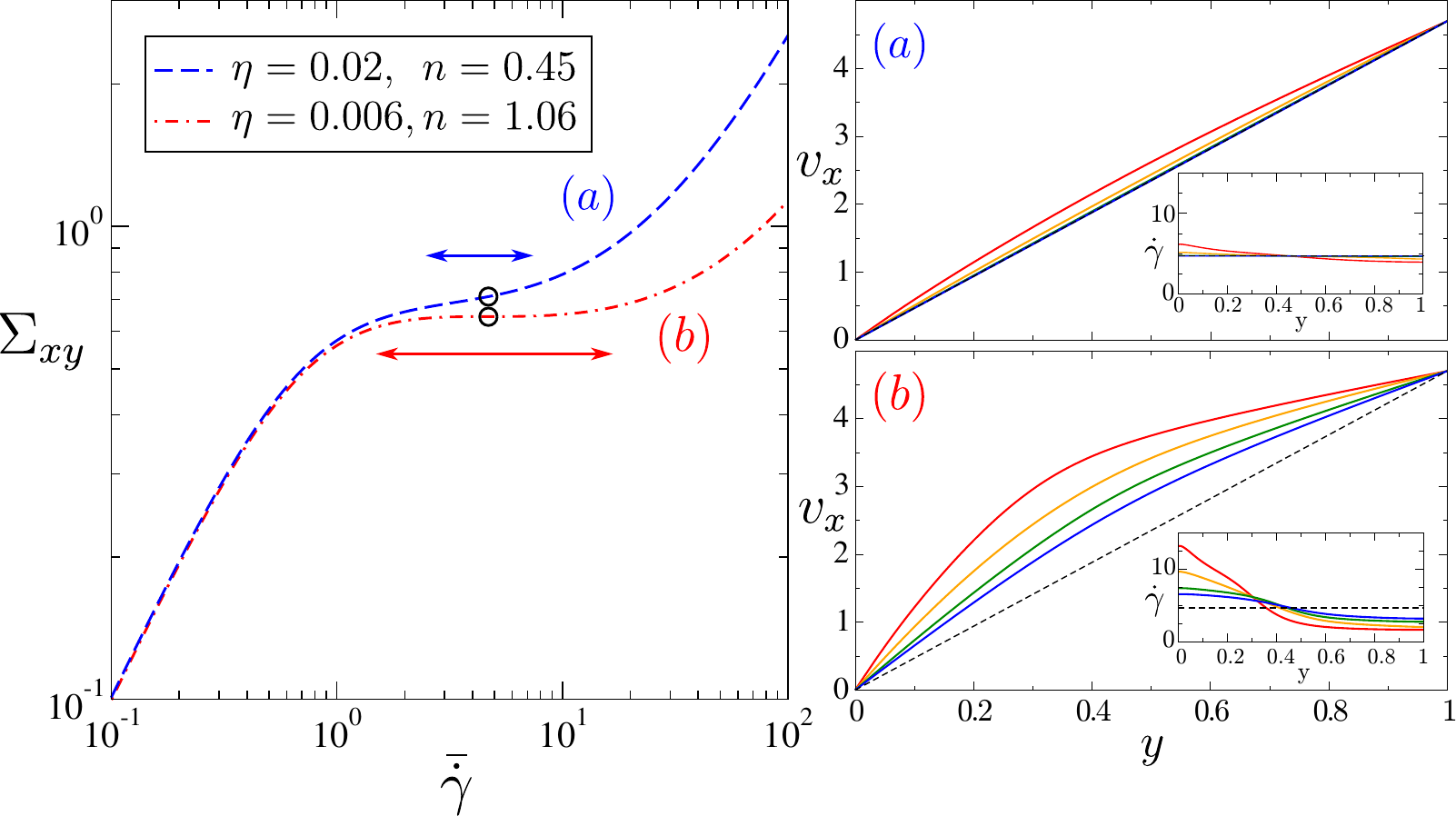}
  \caption{\textbf{Left}: Constitutive curves for (a) moderately and (b) strongly shear thinning fluids, with Newtonian viscosities $\eta = 0.02$ and $0.006$ giving plateau widths $n=0.45$ (blue arrow) and $1.06$ (red arrow) respectively. The shear rate to which the snapshots in Fig.~\ref{fig:snapshots} correspond is shown by the circles. \textbf{Right}: Plots of the velocity profiles (and, inset, shear rate profiles) pertaining to the snapshots of Fig.~\ref{fig:snapshots} at the depths $z = 1, 2, 4, 8L_y$ into the sample from the free surface, for $\eta = 0.02$ $(n=0.45)$ (top) and $\eta = 0.006$ $(n = 1.06)$ (bottom).} \label{fig:constit}
\end{figure}
stress in an imposed flow. The next capture relaxation on a timescale
$\tau$ back towards an unstressed state, with $\alpha$ characterising
the apparent change in relaxation rate as the chains become
anisotropically aligned in flow. The final diffusive term ensures that
the structure of the interface between any shear bands that form is
properly accounted for~\cite{Lu2000}. To test that our results are robust to choice
of constitutive model, we have verified that the physical picture
reported below also holds in the diffusive Johnson-Segalman
model~\cite{Johnson1977} (results not shown).

The air-fluid coexistence is captured via a phase field (Cahn-Hilliard)
approach~\cite{Anderson1998,Kusumaatmaja2016}, with a
mobility $M$ for air-fluid intermolecular diffusion, a scale $G_\mu$
for the free energy density of demixing, and a slightly diffuse
air-fluid interface of thickness $l$. This gives an interfacial
tension $\Gamma=2\sqrt{2}G_\mu l/3$.  In having a diffuse interface,
our simulations are capable of capturing any motion of the contact line
along the wall that arises in flow~\cite{Kusumaatmaja2016}. Our
numerical scheme is described in Ref.~\cite{Hemingway2017}.

We choose units of length in which the gap width $L_y=1$, of time in
which the viscoelastic relaxation time $\tau=1$, and of stress in
which the viscoelastic modulus $G=1$. We set the equilibrium contact
angle of the fluid-air interface at the plates $\theta=90^\circ$, and
have checked that our findings are robust to variations in this
quantity.  We set the inverse air gap size $L_y/(L_z-\Lambda)=0.25$;
the air-fluid interface width $l/L_y=0.01$; the inverse mobility for
intermolecular diffusion, $l^2/MG_\mu
\tau=0.01-0.1$; the stress diffusivity $D=10^{-4}$; and the air viscosity $\eta_a/G\tau=0.006-0.02$: all converged to their appropriate small limit, along with the numerical grid and timesteps.

Important physical quantities to be explored are then the
dimensionless surface tension $\Gamma/GL_y=\Gamma$, sample aspect
ratio $\Lambda/L_y=\Lambda$, Newtonian viscosity $\eta/G\tau=\eta$,
and imposed shear rate $\gdotb\tau=\gdotb$. Among these, we vary the
viscosity $\eta$ in order to vary the shape of the underlying
constitutive curve $\sigma(\gdot)$, for a fixed value of the
anisotropy parameter $\alpha=0.8$. (We could instead have fixed $\eta$
and varied $\alpha$, and have checked that this gives the same
physical picture as that reported below.) In particular, the width of
the plateau region of the constitutive curve will prove an important
quantity in what follows. Accordingly, we define the extrema of the plateau region as the
shear rates $\gdot_h$, $\gdot_l$ that correspond to $\pm 5\%$ of the
stress at the flattest point; then quantify the plateau width via
their logarithmic difference $n=\log(\gdot_{\rm h}/\gdot_{\rm l})$.
We shall report our phase diagrams below both in terms of the
viscosity $\eta - \eta_{\rm c}$ (where $\eta_{\rm c}=0.005918$ is the
value below which the constitutive curve is non-monotonic), and $n$:
the latter is directly set by the former (for our fixed $\alpha$), and
is the more directly accessible quantity experimentally.

\begin{figure}[t!]
  \includegraphics[width=\columnwidth]{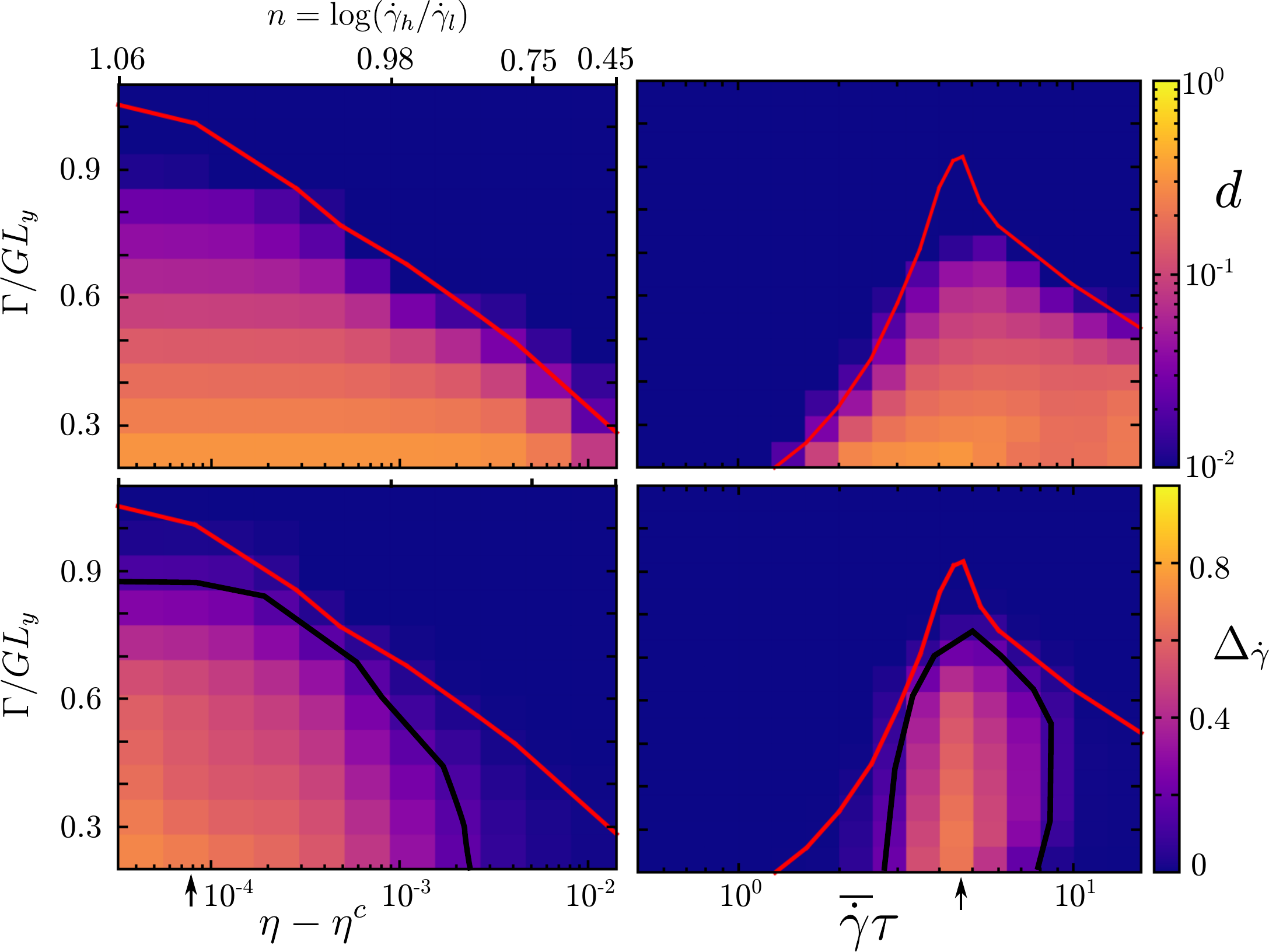}
  \caption{Colourmaps of (\textbf{top}) the degree of bowing $d$ of
  the air-fluid interface and (\textbf{bottom}) the degree of shear
  banding $\dob$ at the cell midpoint $z=8.0$ for a sample of length
  $\Lambda=16.0$. These are shown (\textbf{left}) in the plane of
  surface tension and $\eta - \eta_c$ (bottom x-axis label) or equivalently $n(\eta)$ (top x-axis label,
  characterising the width of the flattest part of the constitutive curve), for a shear
  rate $\gdotb=4.7$ near its flattest part; and (\textbf{right}) in the
  plane of surface tension and imposed shear rate $(\Gamma, \gdotb)$
  for a fixed $\eta=0.006$ ($n=1.06$), which gives a rather flat
constitutive curve. The red lines show the onset of the edge fracture instability, and black lines show the contour $\dobc = 0.15$.}
\label{fig:phasediagrams}
\end{figure}

The surface tension $\Gamma$ and the second normal stress $T_{yy}-T_{zz}=N_2(\gdotb)$, which depends on the imposed shear rate $\gdotb$,
together control the tendency or otherwise of the fluid-air interface
to show edge fracture, as explored in Ref.~\cite{Tanner1983,Hemingway2017}. For a fixed shear rate, the interface is stable at high surface tension
$\Gamma$. At intermediate $\Gamma$, the interface bows modestly but
remains otherwise intact. We define the degree of bowing $d$ as the
difference between the rightmost and leftmost $z-$positions of the
interface, as shown in Fig.~\ref{fig:snapshots}. At low $\Gamma$, full
edge fracture occurs, with a catastrophic interfacial breakup that
would signal the end of any reliable experimental run. Here we focus
on the intermediate regime, with modest edge bowing that is a
precursor to full edge fracture, but might well (in itself) go unnoticed
experimentally. Typical orders of magnitude of $\Gamma=\Gamma/GL_y$
are $0.001-0.1$ for synthetic polymers and $0.1-10.0$
for DNA
solutions~\cite{Schweizer2007,Ricci1984,Boukany2009,Hu2007,Hu2010,Skorski2011}.

We now present our results. The basic phenomenon that we report is
exemplified by the snapshots of Fig.~\ref{fig:snapshots}. Only modest
bowing of the fluid-air interface is apparent in each case, consistent
with the preceding remark. However, radically different bulk behaviour
is seen between the two snapshots. This can be explained by the
differing shape of their underlying constitutive curves. The upper
snapshot pertains to the moderately sloping constitutive curve (a) of
Fig.~\ref{fig:constit}, left. In this case, the disturbance at the
sample edge has virtually no effect on the bulk. In contrast, the
lower snapshot pertains to the flatter constitutive curve (b) in
Fig.~\ref{fig:constit}. Here, any perturbations caused by the modest
disturbance at the sample edge are strongly amplified by the
flatness of the constitutive curve to cause a strong shear banding
effect in the shear rate $\gdotf=\sqrt{2\tens{D}:\tens{D}}$, which invades far into the bulk,
many gap widths in from the sample edge. Crucially, this is true
despite the constitutive curve being monotonically increasing, precluding
true bulk shear banding in the absence of edge effects.

To further exemplify this behaviour, we show in Fig.~\ref{fig:constit} (right) the velocity profiles $v_x(y)$ measured at depths $z = 1, 2, 4, 8\, L_y$ into the sample from the fluid-air interface. For the snapshot of Fig.~\ref{fig:snapshots} (top), these all show near homogeneous shear, with the local shear rate $\gdot=\partial v_x/\partial y$ independent of $y$, apart from some weak heterogeneity in the profile very close to the sample edge, $z=L_y$. In contrast, for the lower snapshot of Fig.~\ref{fig:snapshots} the velocity profiles show noticeable shear banding that persists many gap widths in from the sample edge.  For use below, we note that the `degree of banding' can be quantified for any such profile as $\dob = (\gdot_\textrm{max} -\gdot_\textrm{min})/\gdotb$, with $\gdot_\textrm{max}$ the maximum shear rate at any point across $y$ (which occurs at $y=0$ in Fig.~\ref{fig:constit}, right), and $\gdot_\textrm{min}$ the minimum (which occurs at $y=1.0$). $\gdotbar$ is the gap-averaged value. By inspecting many profiles, we conservatively adopt $\dobc = 0.15$ as the minimum threshold for visually obvious banding.

So far, we have presented results for a single value of the surface
tension $\Gamma$, separately for a moderately sloping constitutive
curve (larger $\eta$) and a flatter curve (smaller $\eta$), with an
imposed shear rate $\gdotb$ near the flattest part of the constitutive
curve in each case. We now explore more fully the behaviour as a
function of $\Gamma,\gdotb$ and $\eta$. To do so, we present in
Fig.~\ref{fig:phasediagrams} (bottom) colourscales of the degree of
banding $\dob$ at the cell midpoint $z=8.0$ for a sample of length
$\Lambda=16.0$. The right panel shows results for $\eta=0.006$ (which
we recall gives the flatter constitutive curve (b) with plateau width
$n=1.06$ in Fig.~\ref{fig:constit}, left) in the plane $\Gamma,\gdotb$
of surface tension and imposed shear rate. (So this panel explores a
range of shear rates across one particular constitutive curve.) The
left panel shows results in the plane $\Gamma,n$ of surface tension
and the parameter $n(\eta)$ that characterises the shape of the
constitutive curve, for a fixed imposed shear $\gdotb=4.7$ near the
flattest part of the constitutive curve in each case. (So this panel
explores a collection of constitutive curves with increasingly broad
plateau regions for increasing $n$ leftwards along the horizontal
axis.)  The corresponding panels in Fig.~\ref{fig:phasediagrams} (top) show the
degree of bowing $d$ of the fluid-air interface, each
directly counterpart to the degree of banding in the panel underneath.

\begin{figure}[t]
  \includegraphics[width=\columnwidth]{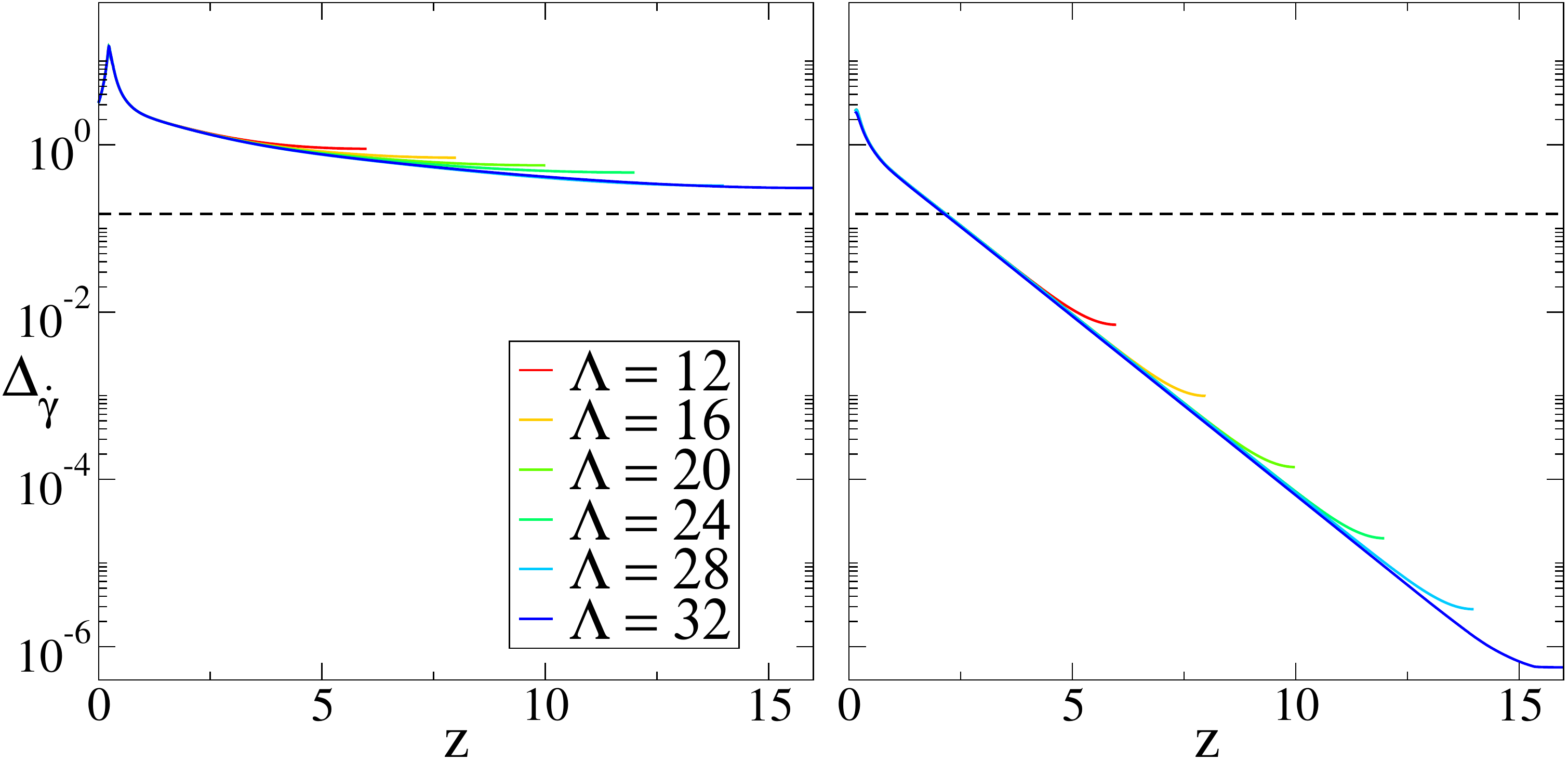}
  \caption{Normalised degree of banding $\dob$ as a function of distance $z$ in from the sample edge, from $z=0$ up to the cell midpoint, for several different sample lengths $\Lambda$. (\textbf{Left:}) for a fluid with a relatively flat constitutive curve, $\eta = 0.006$ ($n=1.06$) and (\textbf{right}) for a fluid with a moderately sloping constitutive curve, $\eta = 0.02$ ($n=0.45$). In each case $\gdotb = 4.7$ and $\Gamma = 0.2$. The horizontal dashed line shows the threshold $\dobc = 0.15$ for visually apparent shear banding.}
  \label{fig:decay}
\end{figure}

The top panels confirm the scenario discussed above from
Ref.~\cite{Hemingway2017}. For any given imposed shear rate $\gdotb$,
the fluid-air interface is undisturbed for high surface tension
$\Gamma$, with zero interfacial bowing, $d=0$. For lower values of the
surface tension, below the red thick line, the interface bows modestly
when the sample is sheared, giving $d=O(L_y)$. (For lower surface
tensions still, not shown in Fig.~\ref{fig:phasediagrams}, full
fracture occurs, giving catastrophic breakup of the interface.)  It is
important to note, however, that the degree of interfacial bowing $d$
does not appear to vary significantly with the overall shape of the
constitutive curve as prescribed by $\eta$ in the top left panel, once
comfortably inside the unstable region.

The degree of shear banding in the plane of surface tension and strain rate in the bottom right panel of Fig.~\ref{fig:phasediagrams} pertains to the flatter constitutive curve (b) of Fig.~\ref{fig:constit}, left. As can be seen, the region of visually apparent banding (as enclosed by the thick black line) arises for shear rates $\gdotbar=2.0-9.0$, in the flattest region of the constitutive curve. For the fixed strain rate $\gdotbar=4.7$ in the flattest part, the degree of banding as a function of surface tension and overall shape of constitutive curve is shown in the bottom left panel of Fig.~\ref{fig:phasediagrams}. A clear relation is seen here between increasing breadth $n$ of the plateau region in the constitutive curve (leftwards along the horizontal axis), and increasing degree of shear banding many gap-widths into the sample. This is true even though the degree of fluid-air interfacial bowing (top left panel) does not vary much with increasing $n$, as emphasized above. This is important, because it shows that strong quasi-bulk shear banding can arise for highly shear thinning fluids, even with a monotonically increasing constitutive curve, even given only modest bowing of the fluid-air interface.

So far, we have presented results for one particular sample length
$\Lambda=16.0$, for the degree of banding at its cell midpoint
$z=8.0$. In Fig.~\ref{fig:decay} we explore the degree of banding as a
function of the position $z$ in from the sample edge, for a range of
different cell sizes. The left panel shows results for the case of the
relatively flat constitutive curve (b) in Fig.~\ref{fig:constit}
(left), and the right panel for the moderately sloping constitutive
curve (a) in Fig.~\ref{fig:constit}. (In each case the imposed shear rate is near the flattest part of the constitutive curve.) As can be seen, for the moderately sloping constitutive curve (Fig.~\ref{fig:decay}, right), the degree of banding falls below the threshold for being visually apparent by a distance of about $2-3$ gap widths in from the sample edge.  In contrast, for the flatter constitutive curve (Fig.~\ref{fig:decay}, left) the degree of banding stays above the threshold for being visually apparent even at the cell centrepoint $z=16.0 L_y$ for the longest sample length $\Lambda=32.0 L_y$. This is  towards the limit of experimental sample aspect ratios, and indeed larger than the depth from the fluid-air surface at which velocimetry is usually performed experimentally.

To summarise, in shear thinning polymeric fluids, we have shown that only modest disturbances of the sample edge (which are the precursors of true edge fracture but might well in themselves go unnoticed experimentally) can lead to strong shear banding that invades far into the fluid bulk, even for the largest sample sizes that are typically studied experimentally. Importantly, this is true even for an underlying constitutive curve that is monotonically increasing, precluding true bulk banding in the absence of edge effects. This work therefore shows that strong quasi-{\em bulk} shear banding can be precipitated by even only modest precursors of the {\em surface} transition of edge fracture.

{\it Acknowledgements -- } The research leading to these results has
received funding from the European Research Council under the EU's 7th
Framework Programme (FP7/2007-2013) / ERC grant number 279365. The
authors thank Mike Cates for a critical reading of the manuscript.

\end{document}